\documentclass[twocolumn]{aastex631}
\renewcommand{\added}{}
\renewcommand{\replaced}[2]{{#2}}
\renewcommand{\deleted}[1]{}
\renewcommand{\edit}[1]{}

\newcommand{\Msol}{\hbox{M$_{\odot}$}}

\newcommand{\hi}{\ion{H}{1}}
\newcommand{\Lsol}{\hbox{L$_\sun$}}

\newcommand{\es}{\hbox{erg~s$^{-1}$}}
\newcommand{\zt}{\mbox{$z\sim2$}}
\newcommand{\CIGALE}{\hbox{\sc cigale}}
\newcommand{\GALFIT}{\hbox{\sc galfit}}

\sfcode`A=1000 \sfcode`B=1000 \sfcode`C=1000 \sfcode`D=1000
\sfcode`E=1000 \sfcode`F=1000 \sfcode`G=1000 \sfcode`H=1000
\sfcode`I=1000 \sfcode`J=1000 \sfcode`K=1000 \sfcode`L=1000
\sfcode`M=1000 \sfcode`N=1000 \sfcode`O=1000 \sfcode`P=1000
\sfcode`Q=1000 \sfcode`R=1000 \sfcode`S=1000 \sfcode`T=1000
\sfcode`U=1000 \sfcode`V=1000 \sfcode`W=1000 \sfcode`X=1000
\sfcode`Y=1000 \sfcode`Z=1000

\begin{document}

\title{A Diverse Population of $z\sim2$ ULIRGs Revealed by JWST Imaging}

\correspondingauthor{S. P. Willner}
\email{swillner@cfa.harvard.edu}

\author[0000-0001-6511-8745]{J.-S. Huang}
\affiliation{Chinese Academy of Sciences South America Center for Astronomy, National Astronomical Observatories, CAS, Beijing, 100101, China}
\affiliation{Center for Astrophysics \textbar\ Harvard \& Smithsonian, 60 Garden St., Cambridge, MA 02138 USA}

\author[0000-0001-7634-1547]{Zi-Jian Li}
\affiliation{Chinese Academy of Sciences South America Center for Astronomy, National Astronomical Observatories, CAS, Beijing, 100101, China}
\affiliation{University of Chinese Academy of Sciences (UCAS), Beijing, 100049, China}

\author[0000-0003-0202-0534]{Cheng Cheng}
\affiliation{Chinese Academy of Sciences South America Center for Astronomy, National Astronomical Observatories, CAS, Beijing, 100101, China}

\author[0000-0001-9062-8309]{Meicun Hou}
\affiliation{The Kavli Institute for Astronomy and Astrophysics, Peking Univerisity, Beijing, China}

\author[0000-0001-7592-7714]{Haojing Yan}
\affiliation{Department of Physics and Astronomy, University of Missouri, Columbia, MO 65211, USA}

\author[0000-0002-9895-5758]{S. P. Willner}
\affiliation{Center for Astrophysics \textbar\ Harvard \& Smithsonian, 60 Garden St., Cambridge, MA 02138 USA}

\author[0000-0002-7928-416X]{Y.-S. Dai}
\affiliation{Chinese Academy of Sciences South America Center for Astronomy, National Astronomical Observatories, CAS, Beijing, 100101, China}

\author[0000-0003-3728-9912]{X. Z. Zheng}
\affiliation{Purple Mountain Observatory, Chinese Academy of Sciences, 10 Yuanhua Road, Qixia District, Nanjing 210023,China}
\affiliation{School of Astronomy and Space Sciences, University of Science and Technology of China, Hefei 230026, China}

\author[0000-0002-4025-7877]{J. Pan}
\affiliation{Chinese Academy of Sciences South America Center for Astronomy, National Astronomical Observatories, CAS, Beijing, 100101, China}
\affiliation{College of Earth Sciences, Guilin University of Technology, Guilin 541004, China}

\author[0000-0001-6854-7545]{D. Rigopoulou}
\affiliation{Department of Astrophysics, Oxford University, Keble Road, Oxford, OX1 3RH, UK}
\affiliation{School of Sciences, European University Cyprus, Diogenes Street, Engomi, 1516 Nicosia, Cyprus}

\author[0000-0002-2504-2421]{T. Wang}
\affiliation{School of Astronomy and Space Science, Nanjing University, Nanjing 210046, China}
\affiliation{Key Laboratory of Modern Astronomy and Astrophysics (Nanjing University), Ministry of Education, Nanjing 210046, China}

\author[0000-0003-0355-6437]{Zhiyuan Li}
\affiliation{School of Astronomy and Space Science, Nanjing University, Nanjing 210046, China}
\affiliation{Key Laboratory of Modern Astronomy and Astrophysics (Nanjing University), Ministry of Education, Nanjing 210046, China}

\author[0000-0001-9143-3781]{Piaoran Liang}
\affiliation{Chinese Academy of Sciences South America Center for Astronomy, National Astronomical Observatories, CAS, Beijing, 100101, China}
\affiliation{University of Chinese Academy of Sciences (UCAS), Beijing, 100049, China}

\author[0000-0003-1845-4900]{A. Esamdin}
\affiliation{Xinjiang Astronomical Observatory, Chinese Academy of Sciences, Urumqi, Xinjiang 830011, China}
\affiliation{University of Chinese Academy of Sciences (UCAS), Beijing, 100049, China}

\author[0000-0002-0670-0708]{G. G. Fazio}
\affiliation{Center for Astrophysics \textbar\ Harvard \& Smithsonian, 60 Garden St., Cambridge, MA 02138 USA}


\begin{abstract}
  \edit1{Four} ultra-luminous infrared galaxies (ULIRGs) observed with JWST/NIRcam in the Cosmos Evolution Early Release Science program \added{offer an unbiased preview of the \zt\ ULIRG population.}  The objects were originally selected at 24~\micron\ and have strong polycyclic aromatic hydrocarbon emission features observed with Spitzer/IRS.   The \edit1{four} objects have similar stellar masses of $\sim$10$^{11}$~\Msol\ but otherwise are quite diverse.  One is  an isolated disk galaxy, but it has an active nucleus as shown by X-ray observations and by a bright point-source \added{nucleus}. \edit1{Two others are merging pairs with mass ratios of 6--7:1. One has} active nuclei in both components, \edit1{while the other has only one active nucleus: the one in the less-massive neighbor, not the ULIRG.} The fourth object is clumpy and irregular and is probably a merger, but there is no sign of an active nucleus. The intrinsic spectral energy distributions for the four AGNs in these systems are typical of type-2 QSOs.  This study is consistent with the idea that even if internal processes can produce large luminosities at $z\sim2$,  galaxy merging may still be necessary for the most luminous objects.
  The diversity of these four initial examples suggests that large samples will be needed to understand the \zt\ ULIRG population.

\end{abstract}



\section{Introduction} \label{sec:intro}

The global star-formation-rate density (GSFRD) of the Universe peaked
about 10~Gyr ago, an epoch that corresponds to redshift $z\sim2$.
This era is often called ``cosmic noon.'' At any redshift, galaxies
of higher mass tend to have higher star-formation rates (SFRs). This
is known as the ``star-formation main sequence.''  Galaxies of all
masses had higher SFR in the past, i.e., the star-formation main
sequence has evolved to lower SFRs over time
\citep{noeske2007,Speagle2014,sandles2022}. \edit1{The $z\sim2$
  population with the highest SFRs, known as ultra-luminous infrared
  galaxies (ULIRGs), was the} predominant contributor to the GSFRD at
cosmic noon \citep{leflock2005,caputi2007, huang2009,
  Eser2014}. Though there are arguments that \edit1{the} SFR at the
high-mass end \edit1{of the star-formation main sequence} might have
been relatively constant rather than continuing to rise with mass
\citep{whitaker2014,leslie2020}, ULIRGs at $z\sim2$ have stellar
masses $M_*>10^{11}$~\Msol\ \citep{huang2009,fang2014}, \edit1{and
  their descendants are today's most massive galaxies}.

The mechanism for triggering the prodigious SFRs characteristic of ULIRGs is unknown, but 
morphological studies can contribute to understanding it. Most local ULIRGs are major mergers 
\citep{sanders1996,bushouse2002,kim2002,sanders2003}, but at higher redshifts, the morphological mix  changes. \citet{kartaltepe2012} studied 52 ULIRGs at $1<z<3$ in the GOODS survey areas and found that only 49\% were mergers. In a study ten times larger in the COSMOS field, \citet{ling2022} found a morphological transition of ULIRG hosts at $z\approx 1.25$, above which merger galaxies dominate and below which disk and merger galaxies contribute similar fractions. At $z\approx 2$, mergers made up $\sim$50\% and disk galaxies  $\sim$22\% of ULIRG hosts. Even if they were a minority, it appears that some galaxies at $z>1$ were so gas-rich that they could  become ULIRGs without merging. More recently, James Webb Space Telescope (JWST) observations that resolve sub-millimeter galaxies (SMGs) in the mid-infrared bands revealed that those qualifying as ULIRGs were all non-merging disk galaxies \citep{cheng2022a, Cheng2022b}.
 
Previous studies of high-$z$ ULIRGs have suffered from two major disadvantages related to severe internal dust extinction. The first disadvantage is that ground-based spectroscopy to determine redshifts is  expensive or even impossible because the dust extinction makes the galaxies so faint. Photometric redshifts remain an option for these galaxies, but the derived redshifts can depend on the amount of extinction \citep{huang2011}, at least for IR-selected galaxies. This leads to large uncertainties in photometric redshifts. The other difficulty is the morphological K-correction \citep{Taylor-Mager2007}. 
Neither 
the stellar population nor the extinction has to be uniform, and therefore galaxies can show different morphologies in different bands.  \edit1{In rest-frame blue light, in particular, a} 
non-uniform distribution of dust can drastically alter a galaxy's \edit1{morphological appearance}, and any older stellar population will be faint.  The longest-wavelength  HST band is F160W, and at $z\ge2$, this band  samples  rest-frame wavelengths  shorter than 5000~\AA, where dusty or old stellar populations are faint or invisible.
 
This paper presents rest-frame near-infrared morphologies of \edit1{four} \zt\ ULIRGs based on
JWST/NIRcam images. The JWST data come from the public Cosmic Evolution Early Release Science Survey (CEERS---\citealt{CEERS3}). The  galaxies  were  identified \edit1{as ULIRGs} by a Spitzer/IRS survey \citep{huang2009, fang2014} and therefore have known redshifts. All four have strong polycyclic-aromatic-hydrocarbon (PAH) emission at rest 7.7~\micron\ and high far-infrared luminosities. \edit1{Three systems} have X-ray detections with the Chandra X-ray Observatory and  presumably contain AGNs. Previous studies have been limited to using spectral-energy-distribution (SED) modeling to separate AGNs and their host galaxies \citep{burgarella2005, noll2009, boquien2019, huang2021}. In contrast, the JWST angular resolution of 0\farcs07--0\farcs14 (=0.6--1.2~kpc at $z=2$)\footnote{For all four systems, the scale is $8.4\pm0.1$~kpc arcsec$^{-1}$.} in the NIRcam 2--4.4~\micron\ bands can show a dusty AGN as a point source distinct from the galaxy disk. This allows us to derive physical parameters for AGNs and their host galaxies based on  morphological separation of the components.
Section~\ref{sec:data} of this paper describes the sample selection, input data, and source SEDs. Section~\ref{sec:merging} discusses source morphologies, and Section~\ref{sec:agn} discusses the AGN components.  Section~\ref{sec:sum} summarizes the results.  All distances are based on a flat $\Lambda$CDM cosmology with $h=0.7$, $\Omega_M=0.272$, and $\Omega_\Lambda=0.728$.

\section{Full SEDs of \zt\ ULIRGs} \label{sec:data}
 
\citet{huang2009} and \citet{fang2014} presented 20--38~\micron\ Spitzer/IRS spectra of  24~$\mu$m-selected sources in the Extended Groth Strip (EGS). The selection criteria, involving 3.6--8.0~\micron\ IRAC colors as well as 24~\micron\ flux density were designed to select star-forming galaxies with strong PAH emission at $1.7\la z\la2.3$ \citep{huang2004,huang2009}. 
Nevertheless, some sample galaxies have an AGN component that produces mid-IR power-law emission \citep{huang2021}. 
Most of the sample galaxies were also detected by  Herschel in the far infrared, and some have been detected at submillimeter and millimeter  wavelengths. All sources have a total infrared luminosity ${>}10^{12}$~\Lsol.
Only \edit1{five} sources from the two IRS samples are within the coverage of the CEERS images. \edit1{One source, EGS~25, is actually at $z=1.35$ \citep{Momcheva2016} rather than $z=1.65$, and  we excluded it from this study.\footnote{\citet{fang2014} misidentified a strong [\ion{Ne}{2}] line at 12.81~\micron\ rest as the 11.3~\micron\ PAH feature. The PAH feature may be present in the IRS spectrum but is weak.}}
The \edit1{remaining four} have redshifts based on strong PAH emission seen in the IRS spectra \citep{huang2009,fang2014}, and the PAH emission (along with the infrared luminosity) is another indicator  of vigorous star formation.
While this initial sample is small, it can be considered a random sample of the $z\approx2$ ULIRG population with \edit1{strong} PAH emission.

\begin{deluxetable*}{lcllcccccc}
%\tabletypesize{\scriptsize}
%\tablewidth{\textwidth} 
\tablenum{1}
\tablecaption{Measured data for the ULIRGs \label{tab:tab1}}
\tablehead{
\colhead{Name} 
& \colhead{[SWM2014]}
& \colhead{RA} 
& \colhead{Dec} 
& \colhead{$F(24~\micron)$}
& \colhead{$z_{\rm IRS}$} 
& \colhead{$z_{\rm 3D}$} 
%& \colhead{$\log_{10} (\frac{M_*^{in}}{\rm M_\odot})$} 
%& \colhead{$A_V^{in}$}
& \colhead{$\log_{10} (\frac{M_*}{\rm M_\odot})$} 
& \colhead{$A_V$}
& \colhead{$\log_{10}(\frac{L_{\rm IR}}{\rm L_\odot})$}\\
&AEGIS
& \multicolumn{2}{c}{J2000}
& {mJy}
}
\startdata 
EGS 11& 23645 
&214.82267 & 52.82264 & 0.59 & 1.80$\pm0.02$ & 1.805 & 11.28$\pm0.10$ & 3.21$\pm0.01$ & 12.54\\
EGS 14&{\it a}
&214.75113 & 52.83003 & 1.05 & 1.87$\pm0.06$ & 1.882 & 11.18$\pm0.09$ & 1.89$\pm0.28$ & 13.00\\
\multicolumn{7}{r}{~~stellar nucleus fit}
& 11.41$\pm0.20$& 2.67$\pm0.01$ \\
\edit1{EGS 22}& {\it b}
&215.16037 & 52.96389 & 0.37 & 1.94$\pm0.10$ & \nodata & 11.29$\pm0.14$ & 2.48$\pm0.47$ & 12.30 \\
EGS 27\tablenotemark{c}&11754
&214.89879 & 52.85250 & 0.49 & 2.29$\pm0.09$ & 2.372 &10.95$\pm0.12$&1.58$\pm0.29$ & 12.56 \\
\multicolumn{7}{r}{~~stellar nucleus fit}& 10.95$\pm0.06$ & 2.14$\pm0.05$\\
\tableline
EGS 14 blue\tablenotemark{d} &\nodata& 214.750720 & 52.829797
&\nodata& \nodata & \nodata&10.55$\pm0.10$&1.11$\pm0.13$ & \nodata\\
\multicolumn{7}{r}{~~stellar nucleus fit} & 10.73$\pm0.14$ & 1.43$\pm0.32$\\
EGS 14 red\tablenotemark{c} &\nodata& 214.751102 & 52.830026 
&\nodata& \nodata& \nodata &11.39$\pm0.18$&2.5$\pm0.5$ & \nodata \\
\multicolumn{7}{r}{~~stellar nucleus fit}& 11.49$\pm0.15$ & 2.49$\pm0.45$\\
\edit1{EGS 22 neighbor\tablenotemark{c}} &\nodata& 215.159845 & 52.963720 
&\nodata& \nodata& \nodata &10.40$\pm0.14$&0.42$\pm0.48$ & \nodata \\
\multicolumn{7}{r}{~~stellar nucleus fit}& 11.02$\pm0.95$ & 1.37$\pm0.54$\\
%EGS 27  & 214.898720 & 52.852445 & \nodata & \nodata &\nodata &\nodata & \nodata& \nodata\\
\enddata
\tablenotetext{a}{blend of 38174 and 38187} 
\tablenotetext{b}{not in 3D-HST area} 
\tablenotetext{c}{X-ray source} 
\tablenotetext{d}{AGN but not an X-ray source} 

\tablecomments{Sources
  above the line represent whole systems. Positions, 24~\micron\ flux
  densities, and IRS redshifts are from \citet{huang2009} for
  EGS~11/14 and \citet{fang2014} for EGS~22/27. EGS~22 has a neighbor 1\farcs3 away which contributes to $\lambda>5.8$~\micron\ photometry. Catalog identifications
  in column~2 are from \citet{Skelton2014}, and 3D-HST redshifts in
  column~7 are from \citet{Momcheva2016}. Sources below the line are
  from this paper. Positions refer to the bright, pointlike nuclei
  and are from the NIRCam F277W images, which give the best
  combination of angular resolution and S/N for the nuclei.  These
  pointlike sources are either AGNs or compact stellar bulges
  surrounding AGNs. \edit1{There are two SED fits for each source, treating the nuclei as AGN or stellar bulges, respectively.}  \added{Physical parameters were derived with \CIGALE\
    and are based on a \citet{Chabrier2003} IMF, delayed exponential star
    formation history, and \citet{Calzetti2000} reddening law.}  }
\end{deluxetable*}

Images used here came from the CEERS image release v0.5 \citep{Bagley2022}\footnote{\url{https://ceers.github.io/dr05.html\#nircam-imaging}} \edit1{for three sources and from newer NIRcam images for EGS~22.
The new images were observed in 2022 December and are not included in CEERS image release v0.5. Therefore, for EGS~22, we utilized the  
MAST\footnote{\url{https://mast.stsci.edu/portal/Mashup/Clients/Mast/Portal.html}} NIRCam stage~3 mosaic calibrated by pipeline v1.8.2 with pmap version \texttt{jwst\_1030.pmap}.}

\begin{figure*}[ht!]
\includegraphics[width=\linewidth]{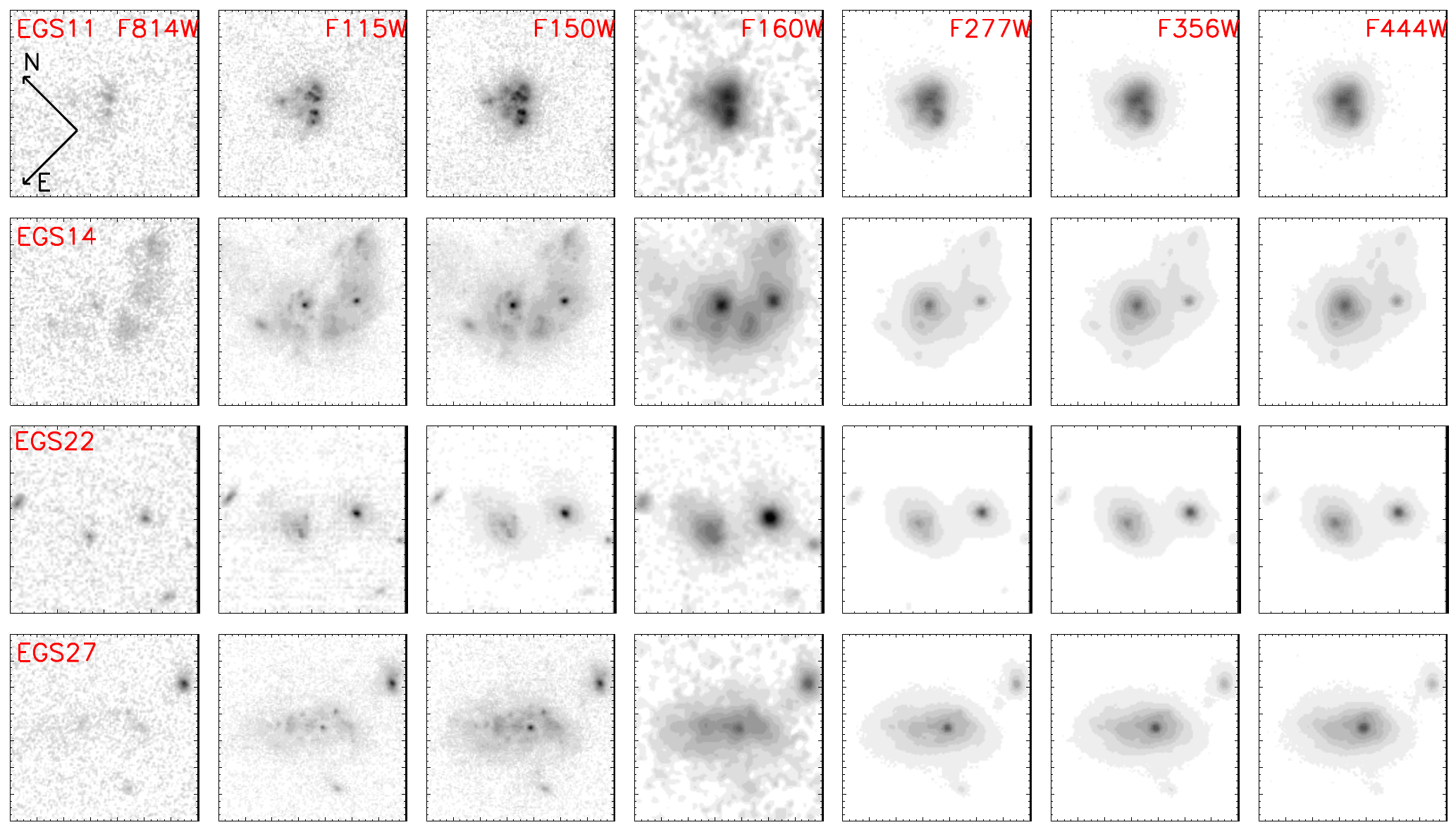}
\caption{Negative stamp images of the four sources.  Each panel is 4\farcs2 in size. Sources are labeled in the left panel of each row. All images have almost the same orientation as shown in the top-left panel.  Images in the leftmost column are HST F814W and in the fourth column HST F160W. Other images are from JWST/NIRCam as labeled in the top row. With only HST images, EGS~14 would be identified as a major merger because the two components have  very similar magnitudes in the HST F160W band.  
\label{fig:fig1}}
\end{figure*}

Figure~\ref{fig:fig1} shows the resulting NIRCam images of the four sources along with existing HST images,
and Table~\ref{tab:tab1} gives the galaxies' properties. 
All four have red colors and show multiple nuclei with surrounding extended structure.   The most complex source is EGS~14, which has two components with differing colors. Much of the southwestern (``blue'') component is seen in the HST F606W (not shown) and F814W images, but the northeastern (``red'') component has only a trace detected in F814W and is invisible in F606W.
\edit1{Both components show} a bright\edit1{, pointlike} nucleus.
Neither nucleus is seen in F606W, and they are faint in F814W.
\edit1{EGS~22 shows two objects in the NIRcam images, 1\farcs3 apart
  (Figure~\ref{fig:fig1}).  The northeastern galaxy shows spiral arms
  with some clumps visible in the short-wave JWST images. A red,
  pointlike nucleus is visible in F277W and longer bands.  The
  southwestern galaxy shows an extended disk around a bright,
  pointlike nucleus. The MIPS 24~\micron\ detection is close to the
  northeastern galaxy, but both may contribute to the detected MIPS
  24~\micron\ flux.  EGS~22 is outside the 3D-HST coverage, so there
  is no other spectroscopic redshift. The IRS spectrum
  \citep{fang2014}  shows only one redshift system. Photometric
  redshifts are 1.83 and 1.70 for the two sources, close to the IRS redshift.  EGS~22 is therefore likely to be a real galaxy pair in the process of merging, as also suggested by a slight distortion of the northeastern galaxy's spiral arms.}
EGS~11 and EGS~27 are barely detected in the \edit1{F606W and F814W} HST bands \added{but obvious in the \edit1{HST 1.6~\micron\ and all} JWST images}. EGS~27 has one bright, pointlike nucleus seen most prominently in the reddest JWST bands \added{and invisible in F814W}.
\added{This nucleus is offset by $\sim$1.4~kpc (in projection) from the centroid of the clumpy galaxy disk.}
EGS~11 has three or more pointlike components, but none of them is as bright as the nuclei seen in the other two sources.

We performed elliptical-aperture photometry with radii matching the 3$\sigma$ surface-brightness contour for each object. 
For EGS~14 \edit1{and EGS~22}, the extended components are well separated from the two nuclei in the F115W and F150W images, and we measured flux densities for the two components separately. The multiple nuclei in EGS~11 show that it is also a merging system, but we treated it as a single galaxy because the components are too close to separate. We also measured flux densities for the point sources in EGS~14 and EGS~27 with 0\farcs09 apertures  corrected to  total magnitudes using the point-spread function in each band. 
\begin{figure*}[ht!]
\centering
\includegraphics[width=0.8\linewidth,clip=true, trim=12 36 0 0]{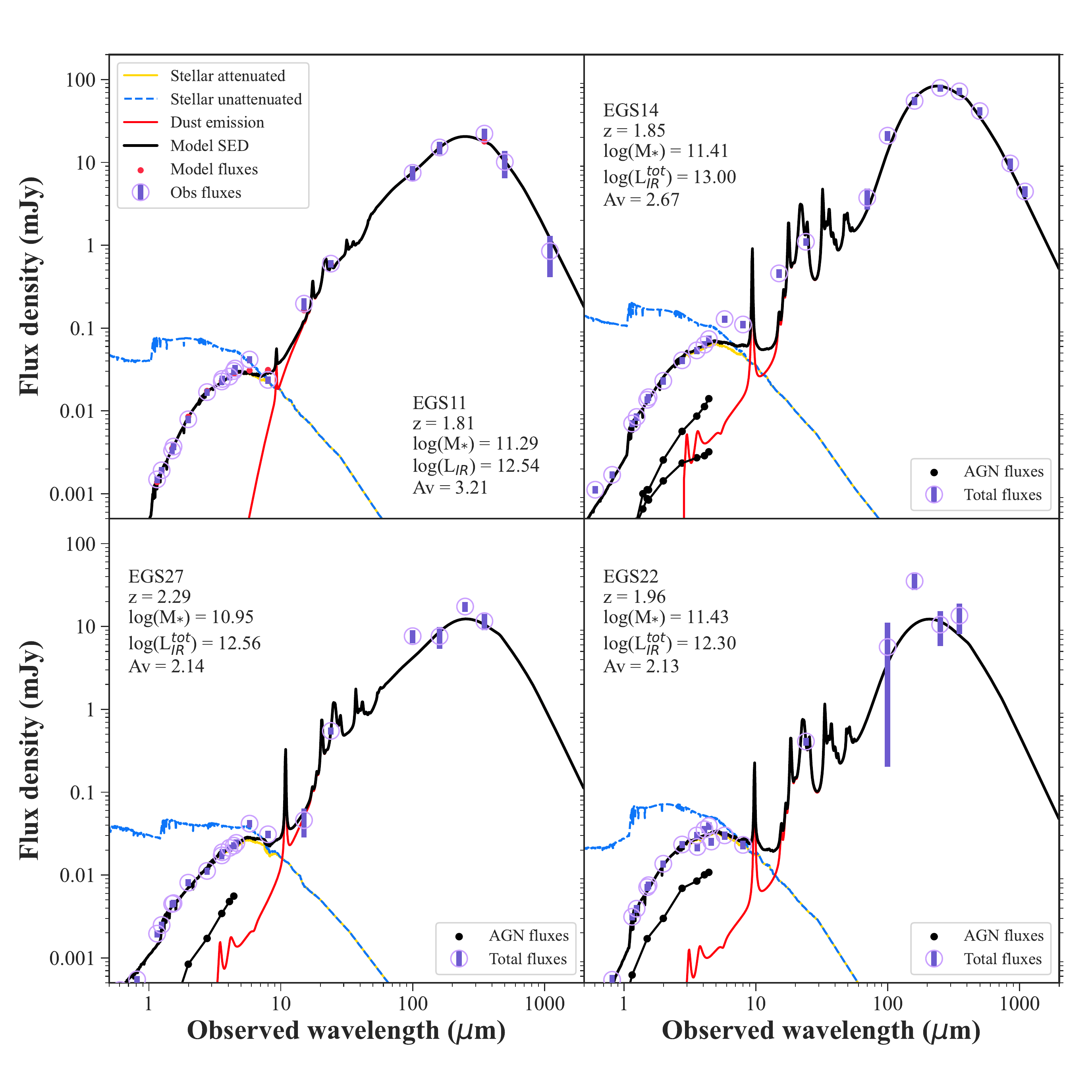}
\caption{SED fits using \CIGALE. Open circles with error bars show the measured flux densities.  \edit1{For EGS~14/22, these are for both components because they cannot be separated at wavelengths beyond 5~\micron. Plain} lines show the model flux densities as indicated in the legend at top left, \edit1{and lines with filled circles show the separate AGN fits. Legends in each panel give fit parameters, which include both components for EGS~14/22.}  The Herschel photometry data are from \citet{lutz2011} and \citet{oliver2012}, and the 
submillimeter and millimeter photometry data are from \citet{zavala2017} and \citet{younger2009}.  The only available constraint on AGN models at wavelength $>$4.44~$\mu$m is that they should not exceed the observed total flux densities.
\label{fig:fig2}}
\end{figure*}

We derived physical parameters for each source by fitting their SEDs with \CIGALE\ \citep{burgarella2005}. Typically \CIGALE\ uses three components---stellar emission, an AGN, and a dusty star-forming component in the infrared bands---to fit the full SED of a galaxy.
The fit yields the percentage of each component's contribution to the total SED. \added{One advantage of SED fitting using JWST images is that we can morphologically separate components,} 
in particular the pointlike components in EGS~14\edit1{/22}/27.  We fitted those \edit1{in two ways. One was to treat the nucleus as an AGN, fit it separately, subtract its light, and then fit the extended component with only stellar emission (and dust extinction).  The other fit treated the nucleus as a stellar bulge and included it as part of the stellar population.}
Whether the pointlike sources are treated as AGNs  or stellar bulges does not significantly change the total stellar masses.
We fitted the blue and red components of EGS~14 \edit1{separately} to determine their  dust extinctions and to obtain the intrinsic SEDs for the pointlike sources in both components. EGS~11 has no bright, pointlike component, \edit1{no X-ray detection, and its 15-to-24~\micron\ flux density ratio is inconsistent with an AGN \citep{huang2009}. If an AGN is present, it isn't contributing to the observed NIRCam fluxes.  We therefore} fit its SED with only stellar and dust components. 
The SEDs and fits are shown in  Figure~\ref{fig:fig2}, and the derived physical parameters are in Table~\ref{tab:tab1}. All galaxies in the four systems have stellar mass  $M_*\sim10^{11}$~\Msol. Fitting the FIR and submillimeter  confirms that all \edit1{systems contain} ULIRGs with EGS~14 close to being a HyperLIRG, \edit1{but the Herschel angular resolution cannot separate the individual galaxies in each system.}
\deleted{For EGS~14, the stellar mass and extinction for the blue and red components are estimated separately and listed in Table~\ref{tab:tab1}.}
%\noindent {\tt\string\documentclass[twocolumn]\{aastex631\}}. \\

%\noindent Note that in the two column style figures and tables will only

%\noindent{\tt\string\begin\{figure*\}} ... {\tt\string\end\{figure*\}}, \\
%\noindent{\tt\string\begin\{table*\}} ... {\tt\string\end\{table*\}}, and \\
%\noindent{\tt\string\begin\{deluxetable*\}} ... {\tt\string\end\{deluxetable*\}}. \\

%\noindent This option is ignored in the onecolumn style.

\section{JWST Morphology of \zt\ ULIRGs} \label{sec:merging}

EGS~11, 14, \edit1{and~22} are merging systems as shown by both HST/WFC3 and JWST/NIRcam images. Each source shows only a single redshift system in Spitzer/IRS and HST/3D spectra (within the uncertainties), further evidence that the components are at least associated. EGS~11 has a clumpy structure with at least a dozen clumps visible in the F150W image. Fewer clumps are visible at longer wavelengths because the angular resolution is worse.  EGS~14 has morphology similar to a familiar merging system, the Antennae, with one component luminous in the rest ultraviolet and the other luminous in the rest near-infrared. The projected distance between the two point-source nuclei is 1\farcs17 = 9.8~kpc,
similar to the 7~kpc separation between the Antennae galaxies. High-redshift, merging ULIRGs resolved with ALMA have typical projected separations of 7--8~kpc \citep{rujopakarn2016, hodge2016,comez2018, rujopakarn2019}, and EGS~14 looks to be another member of this class. \edit1{EGS~22 is also a merging system with a projected separation of 10.9~kpc, larger than those for EGS~11 and~14. It appears to be in an early stage of merging.}

\edit1{In contrast to the other systems,} EGS~27 is an isolated galaxy in the JWST NIRcam bands.  \GALFIT\ confirmed that its morphology in the F444W band is a disk with  S\'ersic index $n=0.7$ and effective radius $r_e=5$~kpc. This is slightly larger than SMGs in SMACS~0723 \citep{cheng2022a} and distinctly larger than $r_e$ of the fainter SMG population \citep{Cheng2022b}.   However, absence of interaction doesn't mean low SFR: EGS~27, despite being an isolated disk, has $\log(L_{\rm IR}/\Lsol) \sim 12.5$, comparable to the two merging sources.

Despite EGS~14's similar appearance to the Antennae, it has 100$\times$ higher infrared luminosity than that system  \citep{sanders2003,seille2022}.
Its normalized SFR, \edit1{$\rm \log(SFR/SFR_{MS})$,} is 0.92, $\gg$0.3~dex above the star-formation main-sequence at $1.5<z<2$ \citep{whitaker2014}, putting this galaxy into the starburst regime \citep{hogan2021}. The normalized SFRs for EGS~11 and EGS~27 are 0.40 and 0.48, respectively, just above the main-sequence upper limit. \edit1{The normalized SFR for EGS~22, using the sum of the two masses, is 0.15.  If all the FIR luminosity comes from the northeastern galaxy, the normalized SFR would be 0.21. These normalized SFRs are high but not significantly above the main sequence. }

At high redshift, merging is more likely in higher-luminosity systems.
An ALMA sample at $z\sim4.5$ with $\log(L_{\rm IR}/\Lsol)\ge12.65$ \citep{comez2018} found that  six of seven systems studied, including all three HyperLIRGs, are minor mergers.\footnote{A major merger requires mass ratio $M_1/M_2<4$---\citealt{man2016}.}  
\edit1{Evidently even a minor merger of two gas-rich galaxies can induce a ULIRG at high redshifts.}
EGS~11/14 have  merging morphologies similar to those of the ALMA sample. The mass ratio for EGS~14 is \replaced{$\sim$5.6}{$\sim$7}, and the mass ratio range for the \citet{comez2018} sample at $z\sim4.5$ is $3<M_{1}/M_2<16$ with only two systems having mass ratios $<$4. \edit1{EGS~22 has a mass ratio $\sim$6, but it has a larger distance
between galaxies than the others, and the system is not in a starburst stage.}
\edit1{The existing data} suggest that a minor merger between two or more gas-rich galaxies  can trigger  luminosities  ${>}10^{13}$~\Lsol.

\section{Identifying AGNs in \zt\ ULIRGs}
\label{sec:agn}

The two bright, pointlike nuclei in EGS~14, \edit1{the one in the EGS~22 neighbor,} and the one in EGS~27 are unresolved in the F115W images. \edit1{(The nucleus of EGS~22 itself is too red to be visible in F115W.)} This implies sizes smaller than 0\farcs1 or 800~pc. The small size suggests the nuclei are AGNs, but based on size alone, they could be compact stellar or gaseous structures.
Stronger evidence is that EGS~14\edit1{/22}/27 were detected by Chandra in 800~ks exposure time. (They are among 25\% of the original PAH-selected samples that have Chandra detections---\citealt{huang2009,fang2014}.) 
In contrast, EGS~11 has no X-ray detection and shows no point source as bright as the ones in EGS~14/22/27. 
The EGS~14 Chandra image shows only one detection at position (214.75116, 52.830010), 0\farcs3 from the red component and 1\farcs4 from the blue component of that system.  The red component is 1.6 magnitudes brighter in the F444W band (rest 1.5~\micron) than the blue component (Figure~\ref{fig:fig2}). Based on both position and near-IR luminosity, the red component is therefore likely to be the dominant contributor to the  X-ray flux. \edit1{The EGS~22 Chandra image also has one detection at position (215.15999, 52.963689).\footnote{\url{https://cxc.cfa.harvard.edu/csc/}
%, \url{https://cxc.cfa.harvard.edu/newsletters/news_28/article6.html}
} 
This is 1\farcs4 from the EGS~22 northeastern galaxy and 0\farcs13 from its southwestern neighbor, which must be the X-ray source. This identification is consistent with this galaxy's bright, pointlike nucleus.} The EGS~27 X-ray emission is unambiguously from the NIRCam point source with an offset between positions of 0\farcs24, well within the Chandra position uncertainty.
 The X-ray luminosities, \added{$L(\hbox{0.5--8~keV})$,} of EGS~14/22/27, even with no correction for \hi\ absorption, are $1.1\times10^{43}$~\es, \edit1{$1.9\times10^{43}$~\es,} and $4.5\times10^{43}$~\es, respectively, well into the AGN range.

\begin{figure}[ht!]
\includegraphics[width=\linewidth]
{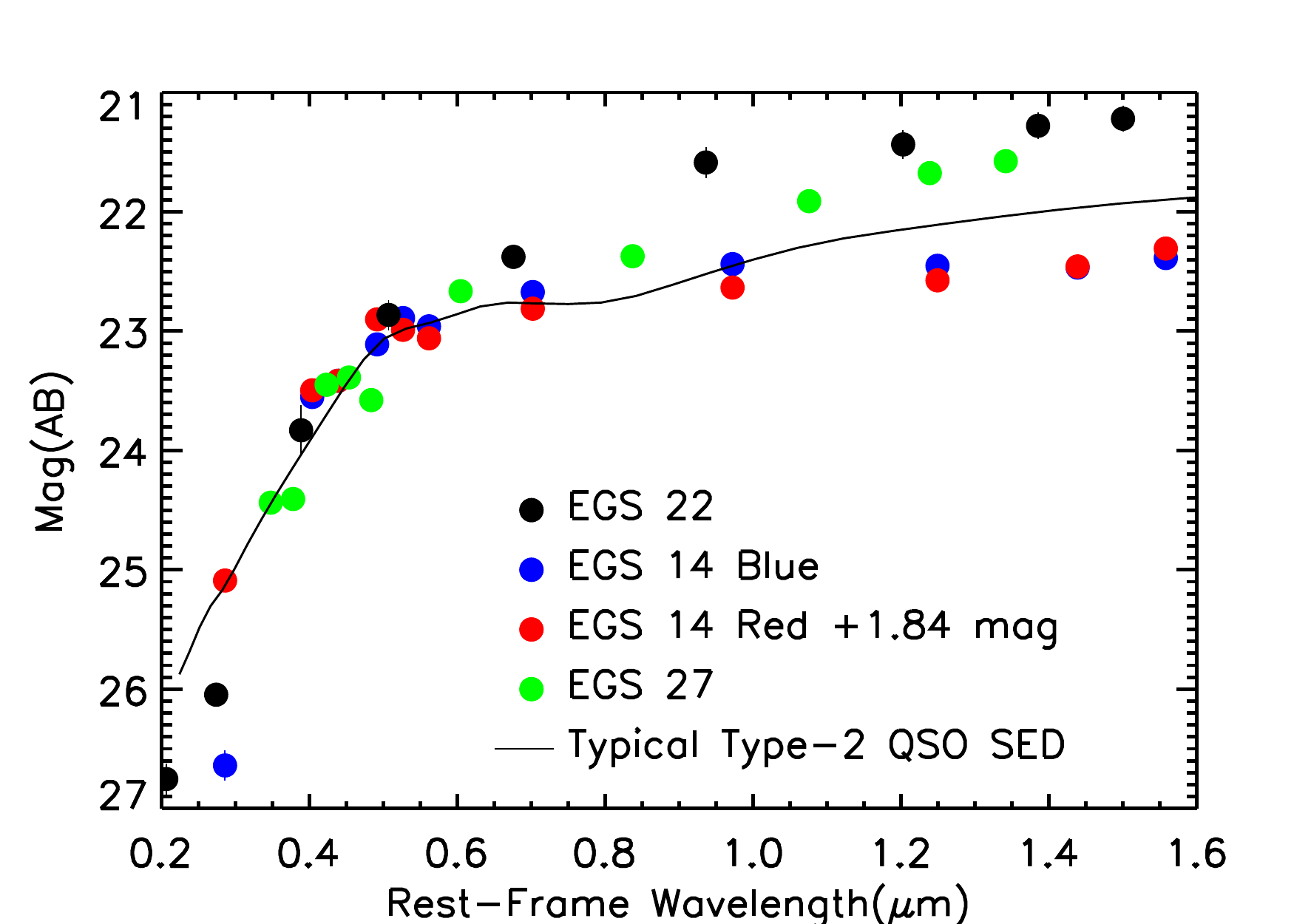}
\caption{SEDs of the four AGN sources identified in Table~\ref{tab:tab1}.
Points show photometry for these sources indicated by color as shown in the legend. The EGS~14 red source is offset (plotted fainter) to allow better comparison with the others.
 All photometry has been corrected for the host-galaxy dust extinction. The solid line is a typical type-2 QSO template \citep{hickox2017} for comparison. All four SEDs show a similar 4000~\AA\ break, as does the template.
\label{fig:fig3}}
\end{figure}

\edit1{EGS~14 has a very different X-ray SED than EGS~22/27 have.} EGS~14's hardness ratio\footnote{${\rm HR}\equiv[(H-S)/(H+S)]$, where $H$ and $S$ are hard (2--8~keV) and soft (0.5--2~keV) counts, respectively.} is 0.32. This implies that it is obscured with a column density possibly larger than 10$^{22}$~cm$^{-2}$. \edit1{EGS~22/27 have similar hardness ratios} of $-$0.29, below the traditional obscured-AGN limit of $-$0.2.  At high redshift, HR is less affected by a given column density, and at $z=1.94$ and $2.31$, \hi\ column densities can be as high as 10$^{22}$~cm$^{-2}$ according to X-ray hardness models \citep{hasinger2008,elvis2012}. In any case, EGS~14 has a column density an order of magnitude higher than the other two X-ray sources, but all the visible--NIR SEDs have a similar shape as shown in Figure~\ref{fig:fig3}.   The host galaxies for the three X-ray sources have  dust extinctions \edit1{in line with the X-ray obscuration} with $A_V=2.5$ for the EGS~14 X-ray counterpart versus $A_V=0.4$ for the EGS~22 neighbor  and $A_V=1.6$ for EGS~27. (These extinctions are for the host galaxies after subtracting the  point sources, as appropriate if the point sources are AGNs.)  The blue point-source in EGS~14 has low dust extinction in its host galaxy, but despite that, no X-ray source was detected. After  correction for host-galaxy extinction, all four point sources have similar SEDs, as shown in Figure~\ref{fig:fig3}, with all four SEDs showing a steep decrease at $\lambda_{\rm rest}\la0.4$~$\mu$m.  A downturn like this is often seen in type-2 QSO SEDs dominated by an old stellar population in the bulge around a central AGN \citep{hickox2017}. The match between the observed SEDs and the type~2 AGN template further confirms that all four point sources are type-2 AGNs. If we ignore the evidence for AGNs and pretend the pointlike sources are  stellar, upper limits for their  masses are $\sim$1.8, 6.3, 13.3, and $4.0\times10^{10}$~\Msol\ for the EGS~14 blue, red, EGS~22 neighbor, and EGS~27 sources, respectively.

\section{Summary}
\label{sec:sum}

Four luminous objects initially selected as Spitzer 24~\micron\ sources have now been observed by JWST/NIRCam.  Redshifts ($1.8\le z\le2.3$) are known from Spitzer/IRS spectra, and luminosities $L_{\rm IR}>10^{12.3}$~\Lsol\ for all systems. \edit1{All four} objects are so red that they are barely detected by HST at wavelengths \edit1{$<$1~\micron}. The NIRcam
images permit morphological studies and measurement of SEDs for each component. One of the four objects is a clumpy disk, probably a merger, with no sign of an AGN.  A second object is a minor-merger system (mass ratio \replaced{5.6}{7}:1) with two AGNs. Evidently even a mass ratio this large can trigger a \edit1{near-}HyperLIRG at $z=1.87$. 
\edit1{The third object is also a likely merger with a neighbor 1\farcs3 away.
The neighbor hosts an AGN.
The merger is likely in an early stage, and its mass ratio is 6:1.} The fourth object is an isolated disk galaxy with an AGN.
Its effective radius is $\sim$5~kpc, about as large as the Milky Way. All four systems have stellar masses near 10$^{11}$~\Msol. The AGN contributions to the broad-band photometry  are less than 25\% and do not significantly change the stellar mass estimates. 

None of the four point sources identified as an AGN was \replaced{detected}{obvious}
in the HST ACS F606W and F814W images. 
In contrast, those objects are easy detections at 2.8~\micron\ and longer wavelengths.  After correction for host-galaxy dust extinction, the four AGNs have SED shapes similar to a typical type-2 QSO. Three of the four AGNs were detected in X-rays by Chandra, but the less-obscured AGN in the EGS~14 merging system was not. The X-ray spectral analysis confirms the AGN nature of the sources and shows high \hi\ column density \edit1{for one of them}.  

The observations give an intriguing first look at the most luminous star-forming galaxies at $z\approx2$. \replaced{, but we need a larger sample to understand the population's properties.  However, even}{If the wide diversity in} these first four examples \replaced{show that the population is quite diverse, and large}{is typical of the population, very large} samples will be needed to understand the ULIRG population at cosmic noon.

\begin{acknowledgments}
This work was sponsored by the National Key R\&D Program of China
grant No.\ 2022YFA1605300, the National Natural Science Foundation
of China (NSFC) grants No.\ 11933003, 12273051, and 11803044, and
in part by the Chinese Academy of Sciences (CAS) through a grant to
the CAS South America Center for Astronomy (CASSACA).
This work is based on observations made with the NASA/ESA/CSA James Webb Space Telescope. 
Data presented in this paper were obtained from the Mikulski Archive for Space
Telescopes at the Space Telescope Science Institute, which is operated by the
Association of Universities for Research in Astronomy, Inc., under NASA contract NAS 5-03127 for JWST. 
These observations are associated with JWST program 1345.
The JWST observations used here can be accessed via \edit1{\url{https://doi.org/10.17909/vdav-5d38}
}.
This research has made use of data obtained from the Chandra Source Catalog, provided by the Chandra X-ray Center (CXC) as part of the Chandra Data Archive.
\end{acknowledgments}


%\vspace{5mm}
\facilities{AKARI, Herschel(PACS, SPIRE), HST(ACS, WFC3-IR), Spitzer(IRAC, MIPS, IRS), JWST(NIRcam), CXO}




%\appendix

%% bibtext sample633
\clearpage
\bibliography{sample631}



\end{document}